# Hailstorm : A Statically-Typed, Purely Functional Language for IoT Applications

Abhiroop Sarkar
sarkara@chalmers.se
Chalmers University
Gothenburg, Sweden

Mary Sheeran
mary.sheeran@chalmers.se
Chalmers University
Gothenburg, Sweden

## ABSTRACT

With the growing ubiquity of *Internet of Things* (IoT), more complex logic is being programmed on resource-constrained IoT devices, almost exclusively using the C programming language. While C provides low-level control over memory, it lacks a number of high-level programming abstractions such as higher-order functions, polymorphism, strong static typing, memory safety, and automatic memory management.

We present Hailstorm, a statically-typed, purely functional programming language that attempts to address the above problem. It is a high-level programming language with a strict typing discipline. It supports features like higher-order functions, tail-recursion, and automatic memory management, to program IoT devices in a declarative manner. Applications running on these devices tend to be heavily dominated by I/O. Hailstorm tracks side effects like I/O in its type system using *resource types*. This choice allowed us to explore the design of a purely functional standalone language, in an area where it is more common to embed a functional core in an imperative shell. The language borrows the combinators of arrowized FRP, but has discrete-time semantics. The design of the full set of combinators is work in progress, driven by examples. So far, we have evaluated Hailstorm by writing standard examples from the literature (earthquake detection, a railway crossing system and various other clocked systems), and also running examples on the GRiSP embedded systems board, through generation of Erlang.

## 1 INTRODUCTION

As the density of IoT devices and diversity in IoT applications continue to increase, both industry and academia are moving towards decentralized system architectures like *edge computing* [38]. In edge computation, devices such as sensors and client applications are provided greater computational power, rather than pushing the data to a backend cloud service for computation. This results in improved response time and saves network bandwidth and energy consumption [50]. In a growing number of applications such as aeronautics and automated vehicles, the real-time computation is more robust and responsive if the edge devices are compute capable.

In a more traditional centralized architecture, the sensors and actuators have little logic in them; they rather act as data relaying services. In such cases, the firmware on the devices is relatively simple and programmed almost exclusively using the C programming language. However with the growing popularity of edge computation, more complex logic is moving to the edge IoT devices. In such circumstances, programs written using C tend to be verbose, error-prone and unsafe [17, 27]. Additionally, IoT applications written in low-level languages are highly prone to security vulnerabilities [7, 58].

Hailstorm is a domain-specific language that attempts to address these issues by bringing ideas and abstractions from the functional and reactive programming communities to programming IoT applications. Hailstorm is a *pure*, *statically-typed* functional programming language. Unlike *impure* functional languages like ML and Scheme, Hailstorm restricts arbitrary side-effects and makes dataflow explicit. The purity and static typing features of the language, aside from providing a preliminary static-analysis tool, provide an essential foundation for embedding advanced language-based security techniques [54] in the future.

The programming model of Hailstorm draws inspiration from the extensive work on Functional Reactive Programming (FRP) [18]. FRP provides an interface to write *reactive* programs such as graphic animations using (1) continuous time-varying values called *Behaviours* and (2) discrete values called *Events*. The original formulation of FRP suffered from a number of shortcomings such as space-leaks [41] and a restrictive form of *stream based I/O*.

A later FRP formulation, *arrowized* FRP [46], fixed space leaks, and in more recent work Winograd-Cort et al. introduced a notion of *resource types* [66] to overcome the shortcomings of the stream based I/O model. The work on resource types is a library in Haskell, and is not suitable to run directly on resource-constrained hardware. Hailstorm uses resource types to uniquely identify each I/O resource. It treats each resource as a signal function to track its lifetime, and prevents resource duplication through the type system. Hailstorm currently has a simple discrete time semantics, though we hope to explore extensions later.

A Hailstorm program is compiled to a dataflow graph, which is executed synchronously. The core of the language is a pure call-by-value implementation of the lambda calculus. The synchronous language of arrowized-FRP provides a minimal set of combinators to which the pure core constructs of Hailstorm can be raised. This language of arrows then enforces a purely functional way to interact with I/O, using resource types.

Hailstorm, in its current version, is a work-in-progress compiler which does not address the reliability concerns associated with node and communication failures plaguing edge devices [58]. We discuss the future extensions of the language to tackle both reliability and security concerns in Section 7. We summarize the contributions of Hailstorm as follows:

- **A statically-typed purely functional language for IoT applications.** Hailstorm provides a tailored, purely functional alternative to the current state of programming resource constrained IoT devices.
- **Resource Types based I/O.** Hailstorm builds on Winograd-Cort et al's work to provide the semantics and implementation of an alternate model of I/O for pure functional languages using *resource types* - which fits the streaming programming model of IoT applications. (Section 4.1)
- **Discrete time implementation** Hailstorm uses the combinators of arrowized FRP in a discrete time setting (Section 3).
- **An implementation of the Hailstorm language.** We implement Hailstorm as a standalone compiler, with Erlang and LLVM backends. We have run case studies on the GRiSP embedded system boards [60], to evaluate the features of the language. (Section 4). The compiler implementation and the examples presented in the paper are made publicly available[1].

## 2 LANGUAGE OVERVIEW

In this section we demonstrate the core features and syntax of Hailstorm using running examples. We start with a simple pure function that computes the $n^{th}$ Fibonacci number.

```
def main : Int = fib 6

def fib : (Int -> Int)
  = fib_help 0 1

def fib_help (a : Int) (b : Int) (n : Int) : Int
  = if n == 0
    then a
    else fib_help (a + b) a (n - 1)
```

The simple program above, besides showing the ML-like syntax of Hailstorm, demonstrates some features like (1) higher-order functions (2) recursion (3) partial application (4) tail-call optimization and (5) static typing. All top-level functions in a Hailstorm program have to be annotated with the types of the arguments and the return type, which currently allows only monomorphic types. However certain built-in combinators supported by the language are polymorphic which will be discussed in the following section.

The pure core of the language only allows writing pure functions which have no form of interactions with the outside world. To introduce I/O and other side effects, we need to describe the concept of a *signal function*.

### 2.1 Signal Functions

A fundamental concept underlying the programming model of Hailstorm is that of a *Signal Function*. Signal Functions, derived from the work on arrowized-FRP [46], are functions that *always* accepts an input and *always* returns an output.

Signal functions are analogous to the nodes of a dataflow graph. Signal functions operate on *signals* which do not have any concrete representation in the language. A signal denotes a discrete value at a give point of time. Nilsson et al [46] use the electric circuit analogy: a signal corresponds to a wire and the current flowing through it, while signal functions correspond to circuit components. An important distinction between Hailstorm and both classic and arrowized-FRP is that signals are always treated as discrete entities in Hailstorm unlike the continuous semantics enforced by FRP.

To create larger programs Hailstorm provides a number of built-in combinators to compose signal functions. These combinators are drawn from the Arrow framework [31] which is a generalization of monads. Arrows allow structuring programs in a *point-free* style, while providing mathematical laws for composition. We start by presenting some of the core Hailstorm combinators[2] and their types for composing signal functions.

```
mapSignal# : (a -> b) -> SF a b
(>>>)      : SF a b -> SF b c -> SF a c
(&&&)      : SF a b -> SF a c -> SF a (b, c)
(***)      : SF a b -> SF c d -> SF (a, c) (b, d)
```

Some of the built-in combinators in Hailstorm are polymorphic and the type parameters a, b and c represent the polymorphic types. mapSignal# is the core combinator which lifts a pure Hailstorm function to the synchronous language of arrows, as a signal function (See Fig 1).

Hailstorm then provides the rest of the built-in combinators to compose signal functions while satisfying nine arrow laws [39]. One of the advantages of having a pure functional language is that such laws can be freely used by an optimizing compiler to aggressively inline and produce optimized code. The semantics of composing signals with the arrow combinators is visually depicted in Fig 1.

---

[1] https://abhiroop.github.io/ppdp-2020-artifact.zip

[2] Non-symbolic built-in combinators & driver functions in Hailstorm end with #

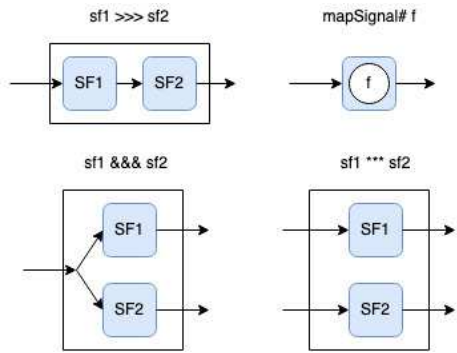

Figure 1: Arrow combinators for signal functions

*Now where do signals actually come from?* To answer this question - a natural extension to signal functions is using them to interact with I/O, which we discuss in the next section.

## 2.2 I/O

Hailstorm adopts a streaming programming model, where an effectful program is constructed by composing various signal functions in the program. The final program is embedded in a stream of input flowing in and the program transforms that into the output.

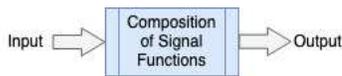

Figure 2: A Hailstorm program interacting with the real world

While this model of I/O adapts well with a pure functional language, it is reminiscent of the now abandoned stream-based I/O interface in Haskell. In Haskell's early stream-based I/O model, the type of the main function was [Response] → [Request]. The major problems with this model are :

- All forms of I/O are restricted to happen at the main function leading to non-modular programs, especially in case of applications running on IoT devices where I/O functions dominate the majority of the program.
- It is non-extensible as the concrete types of Request and Response need to be altered every time any new I/O facility has to be added [48].
- There is also no clear mapping between an individual Request and its Response [48].

Work by Winograd-Cort et al. on resource types [66] attempts to address this problem by *virtualizing* real-world devices as signal functions. What we mean here by "virtualizing" is that the scope of a program is extended so that devices like sensors and actuators are represented using signal functions. For example:

```
sensor   : SF () Int
uart_rx  : SF () Byte
actuator : SF Bool ()
```

We adopt this model in Hailstorm. The type parameter () represents a *void* type. There are no *values* in the language which inhabit the () type. The () type always appears within a signal function. So, an example like sensor : SF () Int - this represents an *action* which when called, produces an integer.

One of the key aspects of designing a pure functional language is this distinction between an expression that returns a *value* and an *action*. When an action (like sensor : SF () Int) is evaluated, it returns a representation of that function call rather than actually *executing* the call itself. This distinction is the key to equational reasoning in a purely functional program. In the absence of such a distinction the following two expressions are no longer equivalent although they represent the same programs.

```
-- Expression 1 : accepts an input and
-- duplicates that input to return a pair
let x = getInput -- makes one I/O call
 in (x,x)

-- Expression 2 : accepts two inputs
-- returns both of them as a pair
(getInput, getInput) -- makes two I/O calls
```

After enforcing a difference between values and action in the language, we soon encounter one of the pitfalls of treating a real-world object as a virtual device - it allows a programmer to write programs with unclear semantics. For example:

```
def foo : SF () (Int, Int) = sensor &&& sensor
```

Although the above program is currently type correct, it can have two conflicting semantics - (1) either sensor &&& sensor implies two consecutive calls to the sensor device or (2) a single call emitting a pair. Given the type of the sensor function, the latter is not supported and the former is incompatible with Hailstorm's discrete, synchronous semantics.

The notion of *Resource Types* seeks to solve this problem by labeling each device with a type-level identifier, such that duplicating a device becomes impossible in the program. We change the type of sensor to :

```
resource S

sensor : SF S () Int
```

The resource keyword in Hailstorm declares a type level identifier which is used for labeling signal functions like sensor above. All the built-in arrow combinators introduced previously are now enriched with new *type-level* rules for composition as follows:

```
mapSignal# : (a -> b) -> SF Empty a b
(>>>) : SF r₁ a b -> SF r₂ b c -> SF (r₁ ∪ r₂) a c
(&&&) : SF r₁ a b -> SF r₂ a c -> SF (r₁ ∪ r₂) a (b, c)
(***) : SF r₁ a b -> SF r₂ c d -> SF (r₁ ∪ r₂) (a, c) (b, d)
```

In the combinators above, the type parameters $r_1, r_2$ represent polymorphic resource type variables, which act as labels for the effectful signal functions. The combinator mapSignal# lifts a pure

Hailstorm function to a signal function without any effectful operations, and as a result the resource type is Empty. The combinators >>>, &&& and ∗ ∗ ∗ compose signal functions, and result in a disjoint-union of the two resources types. This type-level disjoint union prevents us from copying the same resource using any of these combinators. So in Hailstorm if we try this,

```
def foo : SF (S ∪ S) () (Int, Int) = sensor &&& sensor
```

we currently get the following upon compilation:

```
Type-Checking Error:
  Error in "foo":
    Cannot compose resources : S S containing same resource
    Encountered in
       sensor &&& sensor
```

The type rules associated with composing Hailstorm combinators and their operational semantics are presented formally in Section 3.2 and 3.3 respectively.

*2.2.1 Example of performing I/O.* We can distinguish the read and write interface of a resource using two separate resource types. For example, to repeatedly blink an LED we need two APIs - (1) to *read* its status (2) to *write* to it. The drivers for these two functions have the following types:

```
readLed# : SF R () Int
writeLed# : SF W Int ()
```

We use the integer 1 to represent light ON status and 0 for OFF. The program for blinking the LED would be:

```
def main : SF (R ∪ W) () () =
  readLed# >>> mapSignal# flip >>> writedLed#

def flip (s : Int) : Int =
  if (s == 0) then 1 else 0
```

The above program runs the function main infinitely. It is possible to adjust the rate at which we want to run this program, discussed later in Section 2.5. This treatment of I/O as signal functions has the limitation that each device (as well as their various APIs) has to be statically encoded as a resource type in the program.

## 2.3 State

Hailstorm supports stateful operations on signals using the loop# combinator.

```
loop# : c -> SF Empty (a, c) (b, c) -> SF Empty a b
```

The type of the loop# combinator is slightly different from the type provided by the ArrowLoop typeclass in Haskell, in that it allows initializing the state type variable c. The internal body of the signal function encapsulates a polymorphic state entity. This entity is repeatedly fed back as an additional input, upon completion of a whole step of signal processing by the entire dataflow graph. Fig 3 represents the combinator visually.

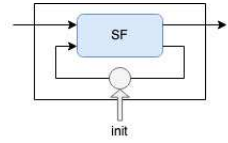

Figure 3: The stateful loop# function

The loop# combinator can be used to construct the delay function as found in synchronous languages like Lustre [26], for encoding state.

```
def delay (x : Int) : SF Empty Int Int =
  loop# x (mapSignal# swap)

def swap (a : Int, s : Int) : (Int, Int) = (s, a)
```

## 2.4 A sample application

We now demonstrate the use of the Hailstorm combinators in a sample application. The application that we choose is a simplified version of an earthquake detection algorithm [65] which was first used by Mainland et al. to demonstrate their domain specific language for wireless sensor networks [42]. The figure below shows the core dataflow graph of the algorithm.

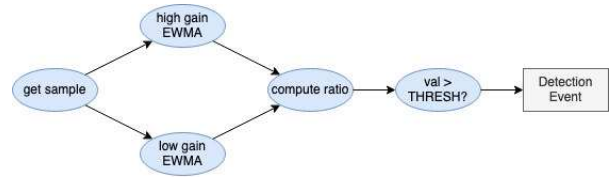

Figure 4: The earthquake detection dataflow graph

The exponentially weighted moving average (EWMA) component above is a stateful element. We assume the getSample input function is a wrapper around a seismometer providing readings of discrete samples. At the rightmost end, the Detection Event would be another stateful entity which would include some form of an *edge detector*. The entire program for the earthquake detection is given in Fig 5.

The function edge in Fig 5 is a stateful edge detector which generates an action if a boolean signal changes from False to True. To program this, we use an imaginary actuator (like an LED) in a GRiSP board, which would glow red once if the input to it is 1, signalling danger, and otherwise stay green signalling no earthquake.

## 2.5 Sampling rate

The combinators introduced so far execute *instantaneously* using a logical clock. In Hailstorm, one logical time step includes the following actions, in sequence -

- accepting a discrete sample of data from each of its connected input devices
- passing the discrete sample through the dataflow graph
- finally passing a discrete value to the responsible actuator

```
resource S
resource E

def main : SF (S ∪ E) () () =
 getSample >>> detect >>> edge

def detect : SF Empty Float Bool
  =   (ewma high &&& ewma low)
  >>> (mapSignal# (\(hi : Float, lo : Float) =>
                    (hi / lo) > thresh))

def ewma (α : Float) : SF Empty Float Float
  = let func = \(x : Float, x_old : Float) =>
                 let x_new = (α *. x) +. (1.0 -. α) *. x_old
                 in (x_new, x_new)
    in loop# 0.0 (mapSignal# func)

-- constants
def low    : Float = ...
def high   : Float = ...
def thresh : Float = ...

def edge : SF E Bool () =
  loop# False (mapSignal# edgeDetector) >>>
  actuator

def edgeDetector (a : Bool, c : Bool) : (Int, Bool) =
  if (c == False && a == True)
  then (1, a)
  else (0, a)

-- getSample : SF S () Float  - Erlang driver
-- actuator  : SF E Int ()    - Erlang driver
```

Figure 5: The earthquake detection algorithm

A program returning a signal function continuously loops around, streaming in input and executing the above steps, at the speed it takes for the instructions to execute. However under most circumstances we might wish to set a slower rate for the program. The rate# combinator is used for that purpose,

```
rate# : Float -> SF r a b -> SF r a b
```

The first argument to rate# is the length of the wall clock time (in seconds) at which we wish to set the period of sampling input. This helps us establish a relation between the wall clock time and Hailstorm's logical clock. We demonstrate the utility of the *rate#* combinator using a *Stopwatch* example.

*2.5.1 Stopwatch.* We program a hypothetical stopwatch which accepts an input stream of Ints where 1 represents START, 2 represents RESET and 3 represents STOP.

```
def f (g : Float) (a : Int, c : Float) : (Float, Float) =
  let inc = c +. g in
  case a of
      1 ~> (inc, inc);
      2 ~> (0.0,0.0);
      _ ~> (c,c)

def stopwatch (g : Float) : SF Empty Int Float
  = rate# g (loop# 0.0 (mapSignal# (f g)))

def main : SF (I U O) () () =
  input >>> stopwatch 1.0 >>> output
```

In the above program, the rate# combinator uses the argument g to set the sampling rate to 1 second, which in turn fixes the granularity of the stopwatch as 1 second.

*2.5.2 Limitation.* In the current implementation of Hailstorm, an operation like (rate# $t_2$ (rate# $t_1$ sf1)) would result in setting the final sampling rate as $t_2$, overwriting the value of $t_1$. In the programs presented here, we use a single clock, and hence a single sampling rate. Libraries like Yampa [14] provide combinators like *delay* :: $Time \rightarrow a \rightarrow SF\ a\ a$ which allow *oversampling* for dealing with multiple discrete sampling rates. As future work, we hope to adopt oversampling operators for communication among signal functions with different sampling rates.

## 2.6 Switches

A Hailstorm dataflow graph allows a form of *dynamic*, data-driven switching within the graph. It accomplishes this using the switch# combinator:

```
switch# : SF r_1 a b -> (b -> SF r_2 b c) -> SF (r_1 ∪ r_2) a c
```

The first argument to switch# accepts a signal function whose output data is used to switch between the various signal functions. The strict typing of Hailstorm restricts the branches of the switch to be of the same type, including the resource type. The switching dataflow is visually presented in Fig 6.

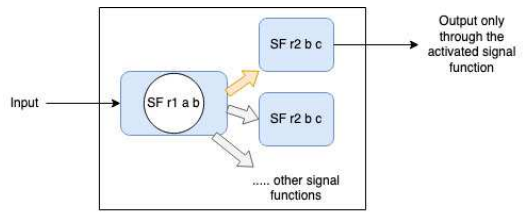

Figure 6: A switch activating only the top signal function

*2.6.1 Limitation.* switch# is a restrictive combinator with a number of known limitations:
- switch# constrains all of its branches to be of the same type. This is particularly restrictive when dealing with actuators where each actuator would have their own resource type. We currently deal with the notion of *choice* in the following way -

```
... >>> foo >>> (actuator1 *** actuator2)
-- instead of emitting a single value foo will emit a pair
-- of values encoded as (move actuator1, dont move actuator2)
```

The Arrow framework provides more useful combinators based on the ArrowChoice typeclass which are currently absent from Hailstorm.

- The switch# combinator is an experiment to describe expressions of the form:

```
switch# input_signal_function
  (\val => if <condition₁ on val>
              then SF₁
           else if <condition₂ on val>
              then SF₂
           else ...)
```

The combinator currently supports expression only of the above form. Thus, we currently do not allow more general functions of type (b→SF b c) as the second parameter to switch#, and so avoid problems with possibly undefined runtime behaviour. However, there is no check in the compiler to enforce this restriction. As future work we hope to adopt the *guards* syntax of Haskell to represent the second parameter as a collection of boolean clauses and their corresponding actions.

## 3 SYNTAX, SEMANTICS AND TYPES

In the previous sections, we have given an informal treatment to the most important syntactic parts of Hailstorm, for IoT applications, and described the programming model using them. In this section, we present the core syntax, type rules and operational semantics of the main parts of Hailstorm.

### 3.1 Syntax

The set of types in the source language is given by the following grammar.

$$\tau ::= () \mid \text{Int} \mid \text{Float} \mid \text{Bool} \mid \tau_1 \to \tau_2 \mid \tau_1 \stackrel{r}{\Longrightarrow} \tau_2 \mid (\tau_1, \tau_2)$$

The type $\tau_1 \stackrel{r}{\Longrightarrow} \tau_2$ represents a signal function type from a to b with the resource type r i.e SF r a b.

The abstract syntax of the core *expressions* of Hailstorm is given by the following grammar. The meta-variable $x \in Var$ ranges over variables of the source language. Additionally we let $i \in \mathbb{Z}$ and $f \in \mathbb{R}$. We use $e$ and $e_{sf}$ separately to denote ordinary expressions and arrow based signal function expressions respectively.

$$
\begin{aligned}
e ::=&\ x \mid \bar{i} \mid \bar{f} \mid \text{True} \mid \text{False} \mid i_1 \text{ binop}_i \ i_2 \mid f_1 \text{ binop}_f \ f_2 \\
&\mid e_1 \text{ relop } e_2 \mid \text{if } e_1 \text{ then } e_2 \text{ else } e_3 \mid \lambda x : \tau. \ e \mid (e_1 \ e_2) \\
&\mid \text{let } x = e_1 \text{ in } e_2 \mid (e_1, e_2) \mid \text{fst\# } e \mid \text{snd\# } e \\
e_{sf} ::=&\ \text{mapSignal\# } (\lambda x.e) \mid e_{sf1} \text{ >>> } e_{sf2} \\
&\mid e_{sf1} \text{ \&\&\& } e_{sf2} \mid e_{sf1} \text{ *** } e_{sf2} \mid \text{loop\# } e_1 \ e_2 \\
&\mid \text{switch\# } e_1 \ (\lambda x.e) \mid \text{read\#} \mid \text{write\#} \\
\text{binop}_i ::=&\ + \mid - \mid * \\
\text{binop}_f ::=&\ +. \mid -. \mid *. \mid / \\
\text{relop} ::=&\ > \mid < \mid >= \mid =< \mid ==
\end{aligned}
$$

In the grammar above we describe two primitives for I/O called read# and write#. In practise, as Hailstorm deals with a number of I/O drivers there exists a variety of I/O primitives with varied parameters and return types. However for the purpose of presenting the operational semantics, we abstract away the complexity of the drivers and use the abstracted read# and write# to describe the semantics in Section 3.3.

### 3.2 Type rules

Hailstorm uses a fairly standard set of type rules except for the notion of *resource types*. The typing context of Hailstorm employs *dual contexts*, in that it maintains (1) $\Gamma$ - a finite map from variables to their types and (2) $\Delta$ - a finite set which tracks all the I/O resources connected to the program.

$$\Delta; \Gamma ::= \cdot \mid \Gamma, x : \tau$$

An empty context is given by ·. Additionally $dom(\Gamma)$ provides the set of variables bound by a typing context.

In Fig 7, we show the most relevant type rules concerning signal functions and their composition. The remaining expressions follow standard set of type rules which is provided in its entirety in the extended version of this paper [55].

Looking at the rule T-Mapsignal, a signal function such as $\tau_1 \stackrel{\emptyset}{\Longrightarrow} \tau_2$ denotes the result of applying mapSignal# to a pure function. This results in an expression with an empty resource type denoted by $\emptyset$. A missing rule is the introduction of a new resource type in the resource type context $\Delta$. The resource type context is an append-only store and a new resource is introduced using the keyword resource. It can be defined using this simple reduction semantics

$$\overline{\Delta; \Gamma \vdash \text{resource } r \rightsquigarrow \Delta \cup r; \Gamma}$$

where $\rightsquigarrow$ denotes one step of reduction which occurs at compile-time. The rules such as T-Compose, T-Fanout, T-Combine, T-Switch apply a type-level *disjoint union* to prevent resource duplication.

### 3.3 Big-step Operational Semantics

In this section, we provide a big-step operational semantics of our implementation of the Hailstorm language, by mapping the meaning of the terms to the lambda calculus. We begin by defining the *values* in lambda calculus that cannot be further reduced:

$$V ::= i \mid f \mid \lambda x.E$$

$i$ and $f$ represent integer and float constants respectively. We use $n$ to represent variables and the last term denotes a lambda expression. The syntax that we use for defining our judgements is of the form :

$$s_1 \vdash M \Downarrow V_i, s_2$$

The variables $s_1, s_2$ are finite partial functions from variables n to their bound values $V_i \in V$. In case a variable $n$ is unbound and $s$ is called with that argument it returns $\emptyset$. The above judgement is read as *starting at state $s_1$ and evaluating the term M results in the irreducible value $V_i \in V$ while setting the final state to $s_2$*.

The first judgement essential for our semantics is this,

$$\overline{s \vdash V \Downarrow V, s}$$

which means that the values V cannot be reduced further.

We use a shorthand notation $\rho(n, s)$ to signify *lookup the variable n in s*. Additionally, to make the semantics more compact we

$$\frac{\Delta;\Gamma \vdash e : \tau_1 \to \tau_2}{\Delta;\Gamma \vdash mapSignal\# \ e : \tau_1 \stackrel{\emptyset}{\Longrightarrow} \tau_2} \text{(T-Mapsignal)}$$

$$\frac{\Delta;\Gamma \vdash e_1 : \tau_1 \stackrel{r_1}{\Longrightarrow} \tau_2 \quad \Delta;\Gamma \vdash e_2 : \tau_2 \stackrel{r_2}{\Longrightarrow} \tau_3 \quad r_1, r_2 \in \Delta \quad r_1 \cap r_2 = \emptyset}{\Delta;\Gamma \vdash e_1 >\!>\!> e_2 : \tau_1 \stackrel{r_1 \cup r_2}{\Longrightarrow} \tau_3} \text{(T-Compose)}$$

$$\frac{\Delta;\Gamma \vdash e_1 : \tau_1 \stackrel{r_1}{\Longrightarrow} \tau_2 \quad \Delta;\Gamma \vdash e_2 : \tau_1 \stackrel{r_2}{\Longrightarrow} \tau_3 \quad r_1, r_2 \in \Delta \quad r_1 \cap r_2 = \emptyset}{\Delta;\Gamma \vdash e_1 \ \&\&\& \ e_2 : \tau_1 \stackrel{r_1 \cup r_2}{\Longrightarrow} (\tau_2, \tau_3)} \text{(T-Fanout)}$$

$$\frac{\Delta;\Gamma \vdash e_1 : \tau_1 \stackrel{r_1}{\Longrightarrow} \tau_2 \quad \Delta;\Gamma \vdash e_2 : \tau_3 \stackrel{r_2}{\Longrightarrow} \tau_4 \quad r_1, r_2 \in \Delta \quad r_1 \cap r_2 = \emptyset}{\Delta;\Gamma \vdash e_1 \ *\!*\!* \ e_2 : (\tau_1, \tau_3) \stackrel{r_1 \cup r_2}{\Longrightarrow} (\tau_2, \tau_4)} \text{(T-Combine)}$$

$$\frac{\Delta;\Gamma \vdash e : (\tau_1, \tau_c) \stackrel{\emptyset}{\Longrightarrow} (\tau_2, \tau_c) \quad \Delta;\Gamma \vdash c : \tau_c}{\Delta;\Gamma \vdash loop\# \ c \ e : \tau_1 \stackrel{\emptyset}{\Longrightarrow} \tau_2} \text{(T-Loop)} \quad \frac{\Delta;\Gamma \vdash t : Float \quad \Delta;\Gamma \vdash e : \tau_1 \stackrel{r}{\Longrightarrow} \tau_2 \quad r \in \Delta}{\Delta;\Gamma \vdash rate\# \ t \ e : \tau_1 \stackrel{r}{\Longrightarrow} \tau_2} \text{(T-Rate)}$$

$$\frac{\Delta;\Gamma \vdash e_1 : \tau_1 \stackrel{r_1}{\Longrightarrow} \tau_2 \quad \Delta;\Gamma \vdash e_2 : \tau_2 \to \tau_2 \stackrel{r_2}{\Longrightarrow} \tau_3 \quad r_1, r_2 \in \Delta \quad r_1 \cap r_2 = \emptyset}{\Delta;\Gamma \vdash switch\# \ e_1 \ e_2 : \tau_1 \stackrel{r_1 \cup r_2}{\Longrightarrow} \tau_3} \text{(T-Switch)}$$

$$\frac{r \in \Delta}{\Delta;\Gamma \vdash read\# : () \stackrel{r}{\Longrightarrow} \tau} \text{(T-Read)} \qquad \frac{r \in \Delta}{\Delta;\Gamma \vdash write\# : \tau \stackrel{r}{\Longrightarrow} ()} \text{(T-Write)}$$

Figure 7: Typing rules of signal functions in Hailstorm

$$\frac{s \vdash exp \Downarrow \lambda x.E, s}{s \vdash mapSignal\# \ exp \Downarrow \lambda x.E, s} \text{(eval-Mapsignal)} \qquad \frac{s_1 \vdash exp_1 \Downarrow V_1, s_2 \quad s_2 \vdash exp_2 \Downarrow V_2, s_3}{s_1 \vdash exp_1 >\!>\!> exp_2 \Downarrow \lambda x.V_2 \ (V_1 \ x), s_3} \text{(eval-Compose)}$$

$$\frac{s_1 \vdash exp_1 \Downarrow V_1, s_2 \quad s_2 \vdash exp_2 \Downarrow V_2, s_3}{s_1 \vdash exp_1 \ \&\&\& \ exp_2 \Downarrow \lambda x. <V_1 \ x, V_2 \ x>, s_3} \text{(eval-Fanout)}$$

$$\frac{s_1 \vdash exp_1 \Downarrow V_1, s_2 \quad s_2 \vdash exp_2 \Downarrow V_2, s_3}{s_1 \vdash exp_1 \ *\!*\!* \ exp_2 \Downarrow \lambda x.\lambda y. <V_1 \ x, V_2 \ y>, s_3} \text{(eval-Combine)}$$

$$\frac{s_1 \vdash init \Downarrow V_i, s_2 \quad s_2 \vdash exp \Downarrow V_1, s_3 \quad n \notin dom(s_3)}{s_1 \vdash loop\#_n \ init \ exp \Downarrow \lambda x.fst \ (V_1 \ (x, V_s)), s_3[n \mapsto snd \ (V_1 \ (x, V_i))]} \text{(eval-Loop-Init)}$$

$$\frac{s_1 \vdash init \Downarrow V_i, s_2 \quad s_2 \vdash exp \Downarrow V_1, s_3 \quad n \in dom(s_3)}{s_1 \vdash loop\#_n \ init \ exp \Downarrow \lambda x.fst \ (V_1 \ (x, \rho(n, s_3))), s_3[n \mapsto snd \ (V_1 \ (x, \rho(n, s_3)))]} \text{(eval-Loop)}$$

$$\frac{s_1 \vdash e_2 \Downarrow V, s_2 \quad s_2 \vdash e_1 \Downarrow \lambda x.E_1, s_3 \quad s_3 \vdash E_1[x \mapsto V] \Downarrow V_f, s_4}{s_1 \vdash (e_1 \ e_2) \Downarrow V_f, s_3} \text{(eval-App)} \quad \frac{s_1 \vdash t \Downarrow V_t, s_2[\Psi \mapsto V_t] \quad s_2[\Psi \mapsto V_t] \vdash exp \Downarrow V, s_3}{s_1 \vdash rate\# \ t \ exp \Downarrow V, s_3} \text{(eval-Rate)}$$

$$\frac{s_1 \vdash exp_1 \Downarrow V_1, s_2 \quad s_2 \vdash exp_2 \Downarrow \lambda b.\sigma_b[\lambda c.E_1, \lambda c.E_2, ...\lambda c.E_n], s_3}{s_1 \vdash switch\# \ exp_1 \ exp_2 \Downarrow \lambda a.((\lambda b.\sigma_b[\lambda c.E_1, \lambda c.E_2, ...\lambda c.E_n])(V_1 \ a)) \ (V_1 \ a), s_3} \text{(eval-Switch)}$$

$$\frac{}{s \vdash read\# \Downarrow \lambda x.read, s} \text{(eval-Read)} \qquad \frac{}{s \vdash write\# \Downarrow \lambda x.(write \ x), s} \text{(eval-Write)}$$

Figure 8: Big-Step Operational Semantics of signal functions in Hailstorm

use pairs <a,b> and their first and second projections, fst, snd. They do not belong to V but it is possible to represent all three of them using plain lambdas and function application - shown in the extended version of this paper [55].

In Fig 8, we show the most relevant big-step operational semantics concerning signal function based combinators. The remaining expressions have standard semantics and the complete rule set is provided in the extended paper [55]. In the rule eval-Rate, we use Ψ to store the sampling rate. In our current implementation, when composing signal functions with different sampling rates, the state transition from $s_2$ to $s_3$ overwrites the first sampling rate.

In eval-Loop-Init and eval-Loop, the subscript n represents a variable name that is used as a key, in the global state map s, to identify each individual state.

For the rule eval-Switch, $\sigma_b[\lambda c.E_1, \lambda c.E_2, ...\lambda c.E_n]$ represents a conditional expression that uses the value of b i.e. ($V_1$ a) to *choose* one of the several branches - $\lambda c.E_i$ - and then supplies ($V_1$ a) again to the selected branch to actually generate a value of the stream.

Of special interest in Fig 8 are the rules eval-Read and eval-Write. We need to extend our lambda calculus based abstract machine with the operations, read and write, to allow any form of I/O. The effectful operations, read and write, are guarded by λs to prevent any further evaluation, and as a result are treated as values. This method is essential to ensure the purity of the language - by treating effectful operations as values.

The program undergoes a *partial evaluation* transformation which evaluates the entire program to get rid of all the $\lambda$s guarding the read operations. Given the expression $\lambda x.read$ the compiler supplies a compile time token of type () which removes the *lambda* and exposes the effectful function read. The partially evaluated program is then prepared to conduct I/O. This approach is detailed further in Section 4.1.

The big-step semantics of the language shows its evaluation strategy. However to understand the streaming, infinite nature of an effectful Hailstorm program we need an additional semantic rule. A Hailstorm function definition is itself an expression and a program is made of a list of such functions,

$$Program ::= main : [Function]$$
$$Function ::= e | e_{sf}$$

Each Hailstorm program compulsorily has a main function. After the entire program is *partially evaluated* (described in Section 4.1) and given that the main function causes a side effect (denoted by () below), we have the following rule:

$$\frac{s_1 \vdash main \Downarrow (), s_2}{s_1 \vdash main \Downarrow main, s_2} \text{(eval-Main)}$$

The eval-Main rule demonstrates the streaming and infinite nature of a Hailstorm program when the main function is a signal function itself. After causing a side effect, it calls itself again and continues the stream of effects while evaluating the program using the semantics of Fig 8.

## 4 IMPLEMENTATION

Here we describe an implementation of the Hailstorm language and programming model presented in the previous sections. We implement the language as a *compiler* - the Hailstorm compiler - whose compilation architecture is described below.

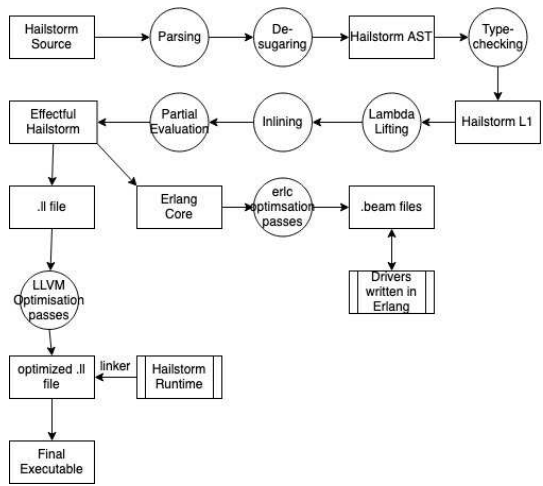

Figure 9: The Hailstorm compilation architecture

The compiler pipeline starts by parsing a Hailstorm source file and *desugaring* small syntactic conveniences provided to reduce code size (such as case expressions). After desugaring, the Hailstorm AST constituted of the grammar described in Section 3.1 is generated. Next, typechecking of the AST using the type rules of Fig 7 is done. After a program is typechecked, it is transformed into the Hailstorm core language called L1.

The L1 language is an enriched version of the simply typed lambda calculus (STLC) with only nine constructors. Unlike STLC, L1 supports recursion by calling the name of a global function. It is currently incapable of recursion using a let binding. L1 has a simpler type system than the Hailstorm source, as it *erases* the notion of resource types - which are exclusively used during type checking.

The L1 language being considerably simpler, forms a sufficient foundation for running correctness preserving optimization passes. L1 attempts to get rid of all closures with free variables, as they are primarily responsible for dynamic memory allocation. Additionally, it attempts to inline expressions (primarily partial applications) to further reduce both stack and heap memory allocation. The optimization passes are described in further detail in Section 4.2.

The final optimization pass in L1 is a *partial evaluation* pass which specializes the program to convert it into an effectful program. Before this pass, all I/O inducing functions are treated as values, by guarding them inside a $\lambda$ abstraction. This pass evaluates those expressions by passing a compile-time token and turning the L1 non-effectful program into an effectful one. This pass is discussed in further detail in Section 4.1.

Finally, the effectful Hailstorm program gets connected to one of its backends. Recursion is individually handled in each of the backends by using a global symbol table. We currently support an LLVM backend [36] and a BEAM backend [4].

### 4.1 I/O

The I/O handling mechanism of the Hailstorm compiler is the essential component in making the language *pure*. A pure functional language allows a programmer to equationally reason about their code. The entire program is written as an *order independent* set of equations. However, to perform I/O in a programming language, it is necessary to (1) enforce an order on the I/O interactions, as they involve *chronological* effects visible to the user and (2) interact with the real world and actually perform an effect.

To solve (1) we use the eval-App rule given in Fig 8. Hailstorm, being a *call-by-value* implementation of the lambda calculus, follows *ordering* in function application by always evaluating the function argument before passing it to the function. This semantics of function application allows us to introduce some form of ordering to the equations.

To solve (2) we have extended our pure lambda calculus core to involve effectful operations like read and write. Again, as observed in the semantic rules, eval-Read and eval-Write from Fig 8, the effectful operations are guarded by $\lambda$ abstractions to treat them as values, rather than operations causing side-effects. This allows us to freely inline or apply any other optimization passes while preserving the correctness of the code. However, to finally perform the side effect, we need to resolve this lambda abstraction at compile time. We do this by *partially evaluating* [33] our program. Below we describe the semantics of the partial evaluation step.

A Hailstorm program is always enforced (by the typechecker in our implementation) to contain a *main* function. We shall address the case of an effectful program i.e a *main* function whose return type is an effectful signal function such as SF r () (). The () type in the input and output parameters reflect that this function reads from a real world source like a sensor and causes an effect such as moving an actuator. The partial application pass is only fired when the program contains the () type in either one of its signal function parameter.

Let us name the main function above with the return type of SF r () () as $main_{sf}$. Now $main_{sf}$ is itself a function, embedded in a stream of input and transforming the input to an output, causing an effect. As it is a function we can write: $main_{sf} = \lambda t.E$.

Additionally in our implementation, after typechecking, a signal function type $\tau_1 \stackrel{r}{\Longrightarrow} \tau_2$ is reduced to a plain arrow type - $\tau_1 \to \tau_2$. Now, using the two aforementioned definitions, we can write the following reduction semantics:

$$\frac{main_{sf} = \lambda t.E \qquad main_{sf} : () \stackrel{r}{\Longrightarrow} () \qquad \tau_1 \stackrel{r}{\Longrightarrow} \tau_2 \rightsquigarrow \tau_1 \to \tau_2}{\frac{\lambda t.E : () \stackrel{r}{\Longrightarrow} () \rightsquigarrow \lambda t.E : () \to ()}{\lambda t.E : () \to () \rightsquigarrow E[t \mapsto \theta : ()]}} \text{(eval-Partial)}$$

where $\rightsquigarrow$ denotes one step of reduction and $\theta$ denotes an arbitrary compile time token of type (). The final step, which produces the expression $E[t \mapsto \theta : ()]$, is the partial evaluation step. We demonstrate this reduction semantics in action using an example:

```
read#  : SF STDIN () Int
write# : SF STDOUT Int ()

def main : SF (STDIN U STDOUT) () () = read# >>> write#
```

We use the eval-Read, eval-Write and eval-Compose rules to translate the *main* function above to

```
main = λt. (λy. write y) ((λx. read) t)
```

Given the above definition of *main*, we can apply the reduction rule eval-Partial and eval-App to get the following,

$$\frac{}{\lambda t. (\lambda y.\ write\ y)\ ((\lambda x.\ read)\ t) \rightsquigarrow_* (\lambda y.\ write\ y)\ (read)} \text{(Partial evaluation)}$$

Now the program is ready to create a *side effect* as the read function is no longer guarded by a $\lambda$ abstraction. The eval-App rule guarantees that read is evaluated first and only then the value is fed to write owing to the *call-by-value* semantics of Hailstorm.

*4.1.1 Limitation.* The Hailstorm type system doesn't prevent a programmer from writing write# >>> read#. The type of such a program would be SF (STDOUT U STDIN) Int Int. As the types do not reflect the () type, the partial evaluation pass is not fired and the program simply generates an unevaluated closure - which is an expected result given the meaningless nature of the program. However, the type would allow composing it with other pure functions and producing bad behaviour if those programs are connected to meaningful I/O functions. This is simply solved by adding the following type rule:

$$\frac{r \in \Delta \qquad r \neq \emptyset \qquad \Delta; \Gamma \vdash \tau_1 = () \lor \tau_2 = ()}{exp : \tau_1 \stackrel{r}{\Longrightarrow} \tau_2} \text{(T-Unsafe)}$$

## 4.2 Optimizations

*4.2.1 Lambda Lifting.* Hailstorm, being a functional language, supports higher order functions (HOFs). HOFs frequently capture free variables which survive the scope of a function call. For example:

```
1  def addFive (nr : Int) : Int =
2    let x = 5 in
3    let addX = \(y:Int) => x + y in
4    addX nr
```

Our implementation treats HOFs as closures which are capable of capturing an environment by allocating the environment on the heap. In line no. 3 above, the value of the variable x is heap allocated. However, in resource constrained devices heap memory allocation, is highly restrictive and any language targeting such devices should attempt to minimize allocation.

Hailstorm applies a *lambda-lifting* [32] transformation to address this. Lambda-lifting lifts a lambda expression with free variables to a top-level function and then updates related call sites with a call to the top-level function. The free variables then act as arguments to the function. This effectively allocates them on the stack (or registers). Owing to our restricted language and the lack of polymorphism, our algorithm is less sophisticated than the original algorithm devised by Johnsson. We describe the operational semantics of our algorithm in the lambda-lifting rule below.

We use a slightly different notation from Section 3.3 here. $P_n$ is used to describe the entire program with its collection of top level functions. We identify a modified program with $P_n[G(x).E]$ to mean a new program with an additional global function G which accepts an argument x and returns an expression E. Finally the $fv$ function is used find the set of free variables in a Hailstorm expression.

$$\frac{P_1 \vdash exp \Downarrow \lambda x.\ E,\ P_2 \qquad fv(\lambda x.\ E) \neq \emptyset}{fv(\lambda x.\ E) = \{i_1, ..., i_n\} \qquad P_2 \vdash \lambda x.\ E \Downarrow E',\ P_3[G(i_1, ..., i_n, x)\ .\ E]}{P_1 \vdash exp \Downarrow E',\ P_3[G(i_1, ..., i_n, x)\ .\ E]}$$

The above rule returns a modified expression $E'$ which consists of a function call to $G(i_1, ..., i_n, x)$ . E with the free variables $i_1, ..., i_n$ as arguments. This rule is repeatedly run on local lambda expressions until none has any free variables. It transforms the program foo above to :

```
def addX' (x:Int) (y:Int) = x + y

def addFive (nr : Int) : Int =
  let x = 5 in
  addX' x nr
```

*4.2.2 Inlining.* The inlining transformation works in tandem with the lambda lifter to reduce memory allocations. Inlining a lambda calculus based language reduces to plain $\beta$ reduction. i.e $((\lambda x.\ M)\ E) \rightsquigarrow M[x \mapsto E]$. Inlining subsumes optimization passes like *copy-propagation* in a functional language. Our prototype inliner is relatively conservative in that (1) it doesn't attempt to inline recursive functions and (2) it doesn't attempt inter-function inlining.

However, it attempts to cooperate with the lambda-lifter to remove all possible sites of partial applications, which are also heap allocated, to minimize memory allocation. The program shown in the previous section, undergoes the cycle in Fig 10

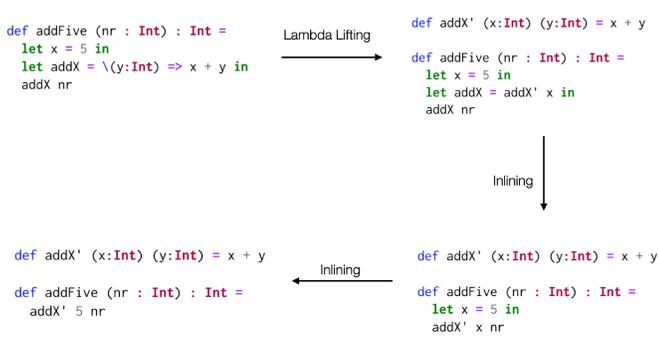

Figure 10: Lambda-lifting and inlining in action

The lambda lifting pass produces a partial application which is further inlined to a single function call where the arguments can be passed using registers. We show the generated LLVM code before and after the optimization passes are run in the extended version of this paper [55]. We show there the absence of any calls to `malloc` in the optimized version of the code. Our inliner doesn't employ any novel techniques but it still has to deal with engineering challenges like dealing with *name capture* [5], which it solves using techniques from the GHC inliner [34].

## 4.3 Code Generation

The Hailstorm compiler is designed as a linear pass through a tower of interpreters which compile away high level features, in the tradition of Reynolds [51]. Here, we show the final code generation for two of the more interesting combinators using C-like notation.

### 4.3.1 loop#.
The `loop#` combinator models a traditional Mealy machine whose output depends on the input as well as the current state of the machine. In the following we see the code generated for the `delay` combinator described in Section 2.3 which is itself described using `loop#`.

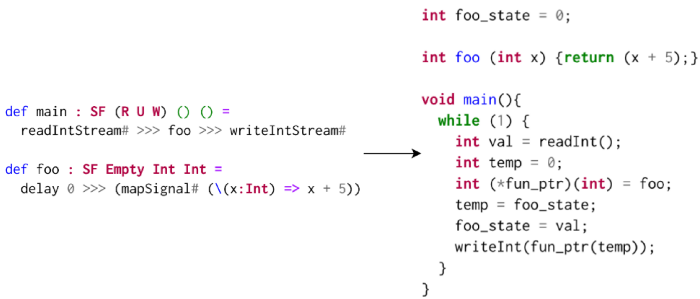

In Erlang, the global variable of `foo_state` is modeled recursively using a global state map, which is updated on every time step. In the LLVM backend, the translation is very similar to the C-notation shown above.

### 4.3.2 switch#.
We show an example of the `switch#` combinator when dealing with stateful branches

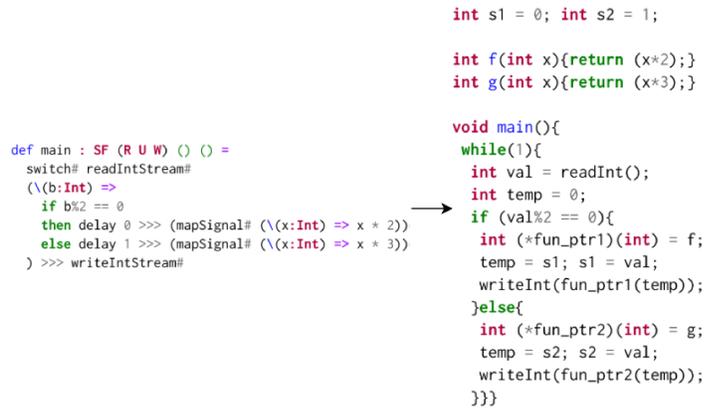

Figure 11: Code generation for switch#

The second parameter in the `switch#` type - (b→SF b c) - acts like a macro which is unfolded into an if-then-else expression in the L1 core language. The core language contains primops for *getting* and *setting* state variables. The final L1 fragment that is generated for the (b→SF b c) component of `switch#` (post *partial evaluation* phase) is given below:

```
... -- s1 and s2 recursively in scope
let x = readInt() in
((λ val .
   if(val % 2 == 0)
   then (λ val2 .
       let temp = s1 in
       let _ = (set s1 to val2) in
       writeInt (f temp))
   else (λ val3 .
       let temp = s2 in
       let _ = (set s2 to val3) in
       writeInt (f temp))
) x) x ....
```

From the above L1 fragment, the C like imperative code shown in Fig. 11 is generated. The if-then-else expressions are translated to case expressions in Erlang and the LLVM translation is very similar to the C code. The global state map now stores two state variables which are updated depending on the value of x. To make the operational aspect of the combinator clearer, we show the evolution of the state variables through the various timesteps of the program:

```
INPUT:  2  3  4  5  6  7  8...
s1:     0  2  2  4  4  6  6...
s2:     1  1  3  3  5  5  7...
OUTPUT: 0  3  4  9  8  15 12...
```

## 4.4 Pull Semantics

In this section we discuss the semantics of data consumption in our compiler. There exist two principle approaches (1) demand-driven pull and (2) data-driven push of data. As Hailstorm's signal function semantics assume a continuous streaming flow of data triggering the dataflow graph, our compiler adopts a pull based approach which continuously *polls* the I/O drivers for data. The program blocks until more data is available from the I/O drivers.

Let us take the earthquake detection example from Section 2.4. The *getSample* input function in the dataflow graph (Fig. 4) is a wrapper around the driver for a simulated seismometer, which when polled for data provides a reading. The rightmost edge of the graph for the `Detection Event` pulls on the dataflow graph after it completes an action and the rest of the graph in turn pulls on `getSample`, which polls the simulated seismometer.

However, in certain devices such as UART, a push based model is more prevalent, where data is asynchronously pushed to the drivers. In such cases, to avoid dropping data the wrapper function (such as `getSample`) needs to be stateful and introduce buffers that store the data. The pseudocode of the wrapper function for such a driver, written for our Erlang backend, would look like Fig 12.

```
loop(State)
  receive
    {hailstormcall, From, Datasize} ->
        (Data, NewState) = extract_Data(Datasize, State),
        From ! {ok, Data},
        loop(NewState);
    {uartdriver, Message} ->
        NewState = buffer(Message, State),
        loop(NewState);
  end.
```

Figure 12: Enforcing pull semantics on push-based data

In Fig 12 there are two separate message calls handled. The data transmission from the drivers is handled using the `uartdriver` message call, which continuously buffers the data. On the other hand, the Hailstorm program, upon finishing one cycle of computation, requests more data using the `hailstormcall` message, and proceeds with the rest of the cycle.

*4.4.1 Limitation.* Elliott has criticized the use of pull semantics [19] as being wasteful in terms of the re-computation required in the dataflow graph. He advocates a hybrid *push-pull* approach, which, in case of continuously changing data, adopts the pull model, but in the absence of any change in the data doesn't trigger the dataflow graph. This approach could be useful in resource constrained devices, where energy consumption is an important parameter, and we hope to experiment with this approach in future work.

## 4.5 The digital - analog interface

The runtime of Hailstorm has to deal with the boundary of discrete digital systems and continuous analog devices. The input drivers have to frequently discretize events that occur at some unknown point of time into a stream of discrete data. An example is the Stopwatch simulation from Section 2.5.1. The pressing of an ON button in a stopwatch translates to the stream of ones (111...) and when switched OFF the simulation treats that as a stream of zeroes. This stream transformation is handled by the wrapper functions around the input drivers.

On the contrary, for the output drivers a reverse translation of discrete to continuous is necessary. We can take the example of operating traffic lights (related demonstration in Section 5.1.2). When operating any particular signal like GREEN supplying a discrete stream of data (even at the lowest granularity) will lead to a *flickering* quality of the light. In that case the wrapper function for the light drivers employs a stateful edge detector, as discussed in Section 2.4, to supply a new signal only in case of change.

## 4.6 Backend specific implementation

The backend implementation includes

- *Memory management.* The compiler attempts to minimize the amount of dynamically allocated memory using lambda-lifting and inlining such that the respective garbage collectors have to work less. Future work hopes to experiment with static memory management schemes like *regions* [61].
- *Tail call optimization (TCO).* Erlang itself does TCO and LLVM supports TCO when using the `fastcc` calling convention.

## 5 EVALUATION

## 5.1 Case Studies

In this section we demonstrate examples from the synchronous language literature [30] written in Hailstorm.

*5.1.1 Watchdog process.* A watchdog process monitors a sequential order processing system. It raises an alarm if processing an order takes more than a threshold time. It has two input signals - (1) *order* : *SF O* () *Bool* which emits *True* when an order is placed and *False* otherwise, (2) *done* : *SF D* () *Bool* which also emits *True* only when an order is done. For output we use - *alarm* : *SF A Bool* () where an alarm is rung only when *True* is supplied. In the program below we keep a threshold time for order processing as 3 seconds.

```
def f ((order : Bool, done : Bool),
       (time : Int, openOrder : Bool)) : (Bool, (Int, Bool))
 = if (openOrder == True && time > 3)
   then (True,(time + 1, False)) -- set alarm once
   else if (done   == True)
        then (False, (0, False)) -- reset
        else if (order   == True)
             then (False, (0, True))
             else (False, (time + 1, openOrder))

def watchdog : SF (O ∪ D ∪ A) () () = (order &&& done) >>>
  (loop# (0, False) (mapSignal# f)) >>> alarm

def main : SF (O ∪ D ∪ A) () () = rate# 1.0 watchdog
```

*5.1.2 A simplified traffic light system.* We take the classic example of a simplified traffic light system from the Lustre literature [53]. The system consists of two traffic lights, governing a junction of two (one-way) streets. In the default case, traffic light 1 is green, traffic light 2 is red. When a car is detected at traffic light 2, the system switches traffic light 1 to red, light 2 to green, waits for 20 seconds, and then switches back to the default situation.

We use a sensor - *sensor* : *SF S* () *Bool* - which keeps returning *True* as long as it detects a car. The system, upon detecting a subsequent car, resets the wait time to another 20 seconds. We sample from the sensor every second. For the traffic lights, we use 1 to indicate green and 0 for red.

```
def lightSwitcher (sig : Bool, time : Int):((Int, Int), Int)
 = if (time > 0)
   then ((0,1), time - 1)
   else if (sig == True)
        then ((0,1), 20) -- reset
        else ((1,0), 0)  -- default

def lightController : SF (S ∪ TL₁ ∪ TL₂) () ((), ()) =
  sensor >>> (loop# 0 (mapSignal# lightSwitcher)) >>>
  (trafficLight₁ *** trafficLight₂)

def main : SF (S ∪ TL₁ ∪ TL₂) () () =
  rate# 1.0 lightController

-- sensor : SF S () Bool
-- trafficLight₁ : SF TL₁ Int ()
-- trafficLight₂ : SF TL₁ Int ()
```

### 5.1.3 A railway level crossing.

*The problem.* We consider a two-track railway level crossing area that is protected by barriers, that must be closed in time on the arrival of a train, on either track. They remain closed until all trains have left the area. The barriers must be closed 30 seconds before the expected time of arrival of a train. When the area becomes free, barriers could be opened, but it's not secure to open them for less than 15 s. So the controller must be warned 45 s before a train arrives. Since the speed of trains may be very different, this speed has to be measured, by detecting the train at two points separated by a known distance. A first detector is placed 2500 m before the crossing, and a second one 100 m after this first. A third is placed after the crossing area, and records a train's leaving. We divide our solution into three programs.

*The Detect Process.* The passage of a train is detected by a mechanical device, producing a *True* pulse - *pulse* : *SF P () Bool* - only when a wheel runs on it (otherwise *False*). The detect process receives all these pulses, but warns the controller only once, on the first wheel. All following pulses are ignored.

```
def f (curr : Bool, old : Bool) : (Bool, Bool) =
  if (curr == True && old == False)
  then (True, curr) else (False, curr)

def detect : SF P () Bool =
  pulse >>> loop# False (mapSignal# f)
```

*A Track Controller.* On each track, a controller receives signals from two detectors. From Detect1 and Detect2, it must compute the train's speed, and warn the barriers 45 seconds before expected time of arrival at crossing. At the maximum speed of 180 km/h, the 100 m between Detect1 and Detect2 are covered in 2 s. So, a clock pulse every 0.1 s would be of good accuracy.

```
def t ((d1 : Bool, d2 : Bool), time:Float) : (Float, Float)
 = case (d1, d2) of
     (True, False) ~> (0.0,0.0);
     (False, True) ~> ((24.0 *. (time +. 0.1) -. 45.0),0.0);
     _             ~> (0.0, time +. 0.1)

def timer : SF Empty (Bool, Bool) Float
 = rate# 0.1 (loop# 0.0 (mapSignal# t))

def trackController : SF (P₁ ∪ P₂) () Float
 = (detect₁ &&& detect₂) >>> timer

def detect₁ : SF P₁ () Bool = ...
def detect₂ : SF P₂ () Bool = ...
```

When a train approaches, the trackController calculates the time for the train to reach the barrier and sends that value. In the absence of a train it sends zeroes.

*The Barriers Controller.* It consists of an alarm - *alarm* : *SF* ($P_1 \cup P_2$) () *Bool* which consumes the time values from the trackController and returns *True* when an alarm is to be rung. The trackController is also sampled every 0.1 second.

```
def g (sig : Float, time : Float) : (Bool, Float) =
  if (sig > 0.0)
  then (False, sig)
  else if (time == 0.1)
       then (True, 0.0)
       else (False, time -. 0.1)

def alarm : SF (P₁ ∪ P₂) () Bool
  = trackController >>> rate# 0.1 (loop# 0.0 (mapSignal# g))
```

Given the alarms from the two separate tracks, the barrier controller sends an open/close signal represented by 0 and 1 respectively. In case a train is approaching in both of the tracks at the same speed - the barrier for only track 1 is opened.

```
def alarm₁ : SF (P₁ ∪ P₂) () Bool = ...
def alarm₂ : SF (P₃ ∪ P₄) () Bool = ...

def openclose (sig : (Bool, Bool)) : (Int, Int) =
  case sig of
    (True, False)  ~> (0, 1);
    (False, True)  ~> (1, 0);
    (True , True)  ~> (0, 1);
    _              ~> (0, 0) -- (False, False)

def barrierController : SF (P₁ ∪ P₂ ∪ P₃ ∪ P₄) () (Int, Int)=
    alarm₁ &&& alarm₂ >>> (mapSignal# openclose)
```

Finally, before sending the signal to the actuators (i.e the barriers), we need an additional system clock that keeps each barrier open for 45 seconds, and ignores other signals in the interim. The handling of the conversion of discrete signals to continuous is done by the drivers for the actuators, as discussed in Section 4.5.

```
def gate ((x : Int, y : Int),
          (t : Float, old : (Int, Int))) :
         ((Int, Int) , (Float, (Int, Int))) =
  if (old ≠ (0,0) ∧ t > 0.0)
  then (old, ((t -. 0.1), old)) -- persisting a signal
  else if (x == 1 ∨ y == 1)
       then ((x,y), (45.0, (x, y)))
       else ((x,y), (0.0, (x,y)))

def main : SF (P₁ ∪ P₂ ∪ P₃ ∪ P₄ ∪ B₁ ∪ B₂) () ((), ()) =
    rate# 0.1 (barrierController >>>
               (loop# (0.0, (0,0)) (mapSignal# gate)) >>>
               (barrier₁ *** barrier₂))
```

```
-- barrier₁ : SF B₁ Int ()
-- barrier₂ : SF B₂ Int ()
```

Note that the three instances of `rate#` all sample at the same interval of 0.1 seconds. Hailstorm currently doesn't have well defined semantics for programs with multiple clock rates.

## 5.2 Microbenchmarks

Here we provide memory consumption and response-time microbenchmarks for the examples presented above using the Erlang backend.

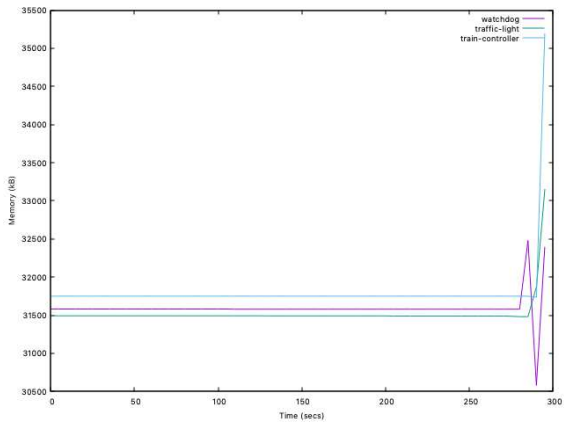

Figure 13: Memory consumption of programs

We measure the mean memory consumption for each program over five runs, each of five minutes duration. Given the I/O driven nature of the programs, the memory consumption shows little to no fluctuations. The Erlang runtime (ERTS) upon initialization sets up the garbage collector, initializes the lookup table and sets up the bytecode-interpreter which occupies 30 MB of memory on average. The actual program and its associated bookkeeping structures takes up an average of 1.5 MB of memory in the programs above. The garbage collector remains inactive throughout the program run. The memory spike visible upon termination is the garbage collector pausing the program and collecting all residual memory.

| Program | Run₁(ms) | Run₂(ms) | Run₃(ms) |
| --- | --- | --- | --- |
| watchdog | 7.7 | 8.65 | 11.4 |
| traffic-light | 3.81 | 3.04 | 2.12 |
| train controller | 29.72 | 28.05 | 29.8 |

Table 1: Response time measured in milliseconds

Table 1 shows the response time for the programs measured in milliseconds. We measure the CPU Kernel time (CPUT) - which calculates the time taken by the dataflow graph to finish one cycle of computation. We show three separate runs where the Erlang virtual machine is killed and restarted to reset the garbage collector. Each of the numbers are an average of twenty iterations of data processing. We use the `erlang:statistics` module for measuring time and in the applications I/O happens over the command line interface, which explains the overall slow behaviour (tens of milliseconds). The metrics are run on the `erts-10.6.4` runtime and virtual machine running on a Macbook-Pro with a 2.9 GHz Intel Core i9 processor. A common observation is that the computation takes less than 1% of the total wall clock time involved in the response rate, showing that the I/O reading/writing times dominate the final response rate.

An alternate benchmarking strategy which we used was to model the input from the sensors as an in-memory structure (in Erlang) and compute the total response time for processing those values using the `timer:tc` module:

| Program | Run₁($\mu s$) | Run₂($\mu s$) | Run₃($\mu s$) |
| --- | --- | --- | --- |
| watchdog | 97.3 | 106.7 | 98.7 |
| traffic-light | 110.8 | 120.2 | 115.1 |
| train controller | 144.3 | 128.1 | 138.7 |

The above values are all measured in microseconds which are averaged over forty iterations each. As expected, using an in-memory structure results in graph processing times that are much lower than those in Table 1.

## 6 RELATED WORK
### 6.1 Programming Languages for IoT

There has been recent work on designing embedded DSLs (EDSLs) for IoT applications [11]. In EDSLs, I/O is handled by embedding a pure core language inside a host language's I/O model - which is an approach that Hailstorm deliberately avoids. Given the I/O dominated nature of IoT apps, we choose to focus much of our attention on designing a composable stream based I/O model, rather than only considering the pure core language, as many EDSLs do.

Other approaches like Velox VM [62] runs general purpose languages like Scheme on specialized virtual machines for IoT devices. A separate line of work has been exploring restrictive, Turing-incomplete, rule-based languages like IoTDSL [2] and CyprIOT [9].

Juniper [29] is one of the few dedicated languages for IoT but it exclusively targets Arduino boards. Emfrp [56] and its successor XFRP [59] are most closely related to the goals of Hailstorm. However, their model of I/O involves writing glue code in C/C++ and embedding the pure functional language inside it. Hailstorm has a more sophisticated I/O integration in the language.

While IoT stands for an umbrella term for a large collection of software areas, there has been research on *declarative* languages for older and *specialized* application areas like:

- *Wireless Sensor Networks (WSNs)*. There exists EDSLs like Flask [42] and macroprogramming languages like Regiment [45] and Kairos [24] for WSNs.
- *Real Time Systems*. Synchronous language like Esterel [10], Lustre [26] are a restrictive set of languages designed specifically for real time systems. Further extensions of these languages like Lucid Synchrone [12], ReactiveML [44], Lucy-n [43] have attempted to makes them more expressive.

The applications demonstrated in the paper are expressible in synchronous languages, albeit using a very different interface from Hailstorm. While languages like Lustre and its extensions are *pure* they restrict their synchronous calculus to the pure core language and handle I/O using the old stream based I/O model of Haskell [48]. Hailstorm explores the design space of *pure* functional programming with the programming model and purity encompassing the I/O parts as well.

In Lustre, a type system called the clock calculus ensures that programs can run without any implicit buffering inside the program. Strong safety properties such as determinism and absence of deadlock are ensured at compile time, and programs are compiled into statically scheduled executable code. This comes at the price of reduced flexibility compared to synchronous dataflow-like systems, particularly in the ease with which bounded buffers can be introduced and used. Mandel et al have studied n-synchronous systems in an attempt to bring greater flexibility to synchronous languages [43]. Hailstorm, in the presence of recursion, is unable to statically predict memory usage but we plan future work on type level encoding of buffer sizes to make memory usage more predictable. *Polychronous* languages like SIGNAL [8] and FRP libraries like Rhine [6] provide static guarantees on correctness of systems with multiple clocks - something that we hope to experiment with in the future.

### 6.2 FRP

Hailstorm draws influence from the FRP programming model. Since the original FRP paper [18], it has seen extensive research over various formulations like arrowized FRP [46], asynchronous FRP [16], higher-order FRP [35], monadic stream functions [47].

Various implementations have explored the choice between a static structure of the dataflow graph (for example Elm [15]) or dynamic structure, as in most Haskell FRP libraries [3] as well as FrTime in Racket [13]. The dynamic graph structure makes the language/library more expressive, allowing programs like sieves [25].

The higher-order FRP implementation offers almost the local maxima of tradeoffs, but at the cost of an extremely sophisticated type system, which infests into the source language, compromising its simplicity.

The loop# combinator in Hailstorm is similar to the $\mu$-combinator first introduced by Sheeran [57] and to the loopB combinator in the Causal Commutative Arrows (CCA) paper [40]. CCA presents a number of mathematical laws on arrows and utilizes them to compile away intermediate structures and generate efficient FRP code - a promising avenue for future work in Hailstorm.

FRP has seen adoption in various application areas. Application areas related to our target areas of IoT applications include robotics [64], real-time systems [63], vehicle simulations [21] and DSLs discussed in detail in the previous section.

### 6.3 Functional I/O

A detailed summary of approaches to functional I/O was presented by Gordon et al. [22]. Since then monadic I/O [23] has become the standard norm for I/O in pure functional languages, with the exception of Clean's I/O system [1] based on uniqueness types.

More recently, there have been attempts at non mondaic I/O using a state passing trick in the Universe framework [20]. The latest work has been on the notion of resource types, proposed by Winograd-Cort et al. [66], which is explored as a library in Haskell. In effect, their library uses Haskell's monadic I/O model; however, they mention possible future work on designing a dedicated language for resource types. Hailstorm explores that possibility by integrating the idea of resource types natively in a language's I/O model. FRPNow! [49] provides an alternate monadic approach to integrate I/O with FRP.

## 7 FUTURE WORK

Hailstorm is an ongoing work to run a pure, statically-typed, functional programming language on memory constrained edge devices. As such, there are a number of open avenues for research:

- *Security.* We hope to integrate support for Information Flow Control (IFC) [28] - which uses language based privacy policies to determine safe dataflow - in Hailstorm. Given that the effectful combinators are based on the Arrow framework, we expect to build on two lines of work - (1) integrating IFC with the Arrow typeclass in Haskell [37] (2) a typeclass based technique to integrate IFC without modifying the compiler [52] but one which relies on the purity and static typing of the language.

- *Reliability.* Hailstorm currently doesn't provide any fault tolerance strategies to mitigate various node/communication failures. However, being hosted on the Erlang backend, we plan to experiment with distributed versions of the arrow combinators where the underlying runtime would use *supervision trees* to handle failures. This would be a much more invasive change as the synchronous dataflow model of the language is not practically suitable in a distributed scenario - leading to more interesting research directions on *macro-programming models* [24].

- *Memory constrained devices.* The evaluation of this paper is carried out on GRiSP boards which are relatively powerful boards. We are currently working on developing a small virtual machine which can interpret a functional bytecode instruction set and run on much more memory constrained devices like STM32 microcontrollers.

## 8 CONCLUSION

We have presented the design and implementation of Hailstorm, a domain specific language targeting IoT applications. Our evaluation suggests that Hailstorm can be used to declaratively program moderately complex applications in a concise and safe manner. In the future, we hope to use the purity and type system of Hailstorm to enforce language-based security constraints, as well as to increase its expressiveness to enable the description of interacting IoT devices in a distributed system.

## ACKNOWLEDGMENTS

This work was funded by the Swedish Foundation for Strategic Research (SSF) under the project Octopi (Ref. RIT17-0023). We would also like to thank Henrik Nilsson and Joel Svensson for their valuable feedback on improving our paper.